\documentstyle[emulateapj,apjfonts,psfig]{article}

\lefthead{J.~M.~Miller et al.}  
\righthead{X-ray and Radio Monitoring of M81*}

\received{Date}
\revised{Date}
\accepted{Date}

\journalid{vol}{date}
\articleid{1}{4}
\paperid{id}

\cpright{AAS}{1999}
\ccc{x}

\begin{document}

\title{Exploring Accretion and Disk-Jet Connections in the LLAGN M81*}

\author{J.~M.~Miller\altaffilmark{1},
     	M.~Nowak\altaffilmark{2},
        S.~Markoff\altaffilmark{3},
        M.~P.~Rupen\altaffilmark{4},
        D.~Maitra\altaffilmark{1}}

\altaffiltext{1}{Department of Astronomy, University of Michigan, 500
Church Street, Ann Arbor, MI 48109, jonmm@umich.edu}

\altaffiltext{2}{MIT Kavli Institute for Astrophysics and Space
Research, 70 Vassar Street, Cambridge, MA, 01239}

\altaffiltext{3}{Science Park 904, 1098 XH, Amsterdam NL}

\altaffiltext{4}{National Radio Astronomy Observatory, 1003 Lopezville
Road, Socorro NM 87801}


\authoremail{jonmm@umich.edu}

\label{firstpage}

\begin{abstract}
We report on a year-long effort to monitor the central supermassive
black hole in M81 in the X-ray and radio bands.  Using {\it Chandra}
and the VLA, we obtained quasi-simultaneous observations of M81* on
seven occasions during 2006.  The X-ray and radio luminosity of M81*
are not strongly correlated on the approximately 20-day sampling
timescale of our observations, which is commensurate with viscous
timescales in the inner flow and orbital timecales in a
radially-truncated disk.  This suggests that short-term variations in
black hole activity may not be rigidly governed by the ``fundamental
plane'', but rather adhere to the plane in a time-averaged sense.
Fits to the X-ray spectra of M81* with bremsstrahlung models give
temperatures that are inconsistent with the outer regions of very
simple advection-dominated inflows.  However, our results are
consistent with the X-ray emission originating in a transition region
where a truncated disk and advective flow may overlap.  We discuss our
results in the context of models for black holes accreting at small
fractions of their Eddington limit, and the fundamental plane of black
hole accretion.
\end{abstract}

\section{Introduction}
Low-luminosity active galactic nuclei (LLAGN) are potentially
important transition objects, harboring supermassive black holes that
accrete at a rate that is intermediate between Seyfert AGN and
quasars, and extremely under-luminous sources such as Sgr A*.  LLAGN
may provide clues to jet production: in these systems, compact
relativistic radio jets are often detected (Nagar et al.\ 2002,
Anderson \& Ulvestad 2005), and the natural time scales are such that
the details of jet production can be revealed.  Moreover, LLAGN are
often ``radio-loud'' (Ho 2008), meaning that jets are an important
part of the overall accretion flow.  At a distance of only 3.6~Mpc
(Freedman et al.\ 1994), the accreting supermassive black hole at the
center of M81 powers the nearest LLAGN, M81*.

The nature of the innermost accretion flow in LLAGN is not clear.  It
is likely that these sources are still fueled partially by an
accretion disk - double-peaked optical emission lines are seen in M81*
(Bower et al.\ 1996; also see Devereux \& Shearer 2007) - but
relativistic X-ray lines from the inner accretion disk are not clearly
detected in these systems (e.g. Dewangan et al.\ 2004, Reynolds et
al.\ 2009; for a review, see Miller 2007).  The inner disk may be
truncated, and the innermost flow may be advection--dominated (Narayan
\& Yi 1994; also see Blandford \& Begelman 1999).  Theoretical work
suggests that thick advective disks and radial flows may help to
maintain poloidal magnetic fields and power jets (e.g. Livio 2000;
Meier 2001; Reynolds, Garofalo, \& Begelman 2006).  X-ray observations
can test and refine models for advective inflow at low mass accretion
rates in a variety of ways.  For instance, the inner accretion flow is
predicted be extremely hot, with temperatures ranging between
$10^{12}$~K centrally to $10^{9-10}$~K in their outermost radii
(Narayan \& Yi 1994, 1995).  Recent observations of X-ray binaries
have achieved the sensitivity required to test these predictions
(e.g. Bradley et al.\ 2007), and find emission consistent with much
lower temperatures.

X-ray emission is often used as a trace of the accretion inflow
(although some X-ray emission could originate in a jet; e.g. Markoff,
Falcke, \& Fender 2001; also see Miller et al.\ 2002, Russell et al.\
2010), and radio emission is used to trace the jet power.  In X-ray
binaries, X-ray and radio emission follows the relationship $L_{R}
\propto {L}_{ X}^{0.7}$, both in ensemble and in individual sources
(Gallo, Fender, \& Pooley 2003; however, see Jonker et al.\ 2009).
This relation has been generalized into a fundamental plane of black
hole accretion, combining radio luminosity, X-ray luminosity, and
black hole mass (Merloni, Heinz, \& Di Matteo 2003; Falcke, Kording,
\& Markoff 2004; Gultekin et al.\ 2009).  If accretion physics scales
predictably with black hole mass, then for any individual object of
known mass, the relationship between radio and X-ray emission should
be fixed, on average.

Prior to discrete jet ejection events in stellar-mass black holes,
quasi-periodic oscillations (QPOs) are often observed in the X-ray
flux, with a characteristic frequency of $\sim6$ Hz (see, e.g.,
Nespoli et al.\ 2003, Ferreira et al.\ 2006, Klein-Wolt \& van der
Klis 2008, Fender, Homan, \& Belloni 2009).  If this frequency is a
Keplerian orbital ferequency, it corresponds to a radius of $66~{\rm
GM}/{\rm c}^{2}$ for a black hole with a mass of $10~{\rm M}_{\odot}$.
This radius is broadly consistent with lower limits on the inner edge
of the accretion disk in M81* based on the width of the Fe K$\alpha$
emission line (Dewangan et al.\ 2004, Young et al.\ 2007).  In stellar
mass systems, it is not possible to test disk-jet connections on the
period defined by the QPO, although the oscillation might be tied to
jet production.  In supermassive black holes, however, this timescale
is accessible.  For a black hole of $7\times 10^{7}~{\rm M}_{\odot}$
like that in M81* (Devereux et al.\ 2003), monitoring every 2--4 weeks
can sample the corresponding timescale.

\centerline{~\psfig{file=f1.ps,width=3.2in,angle=-90}~}
\figcaption[t]{\footnotesize The plot above shows the combined MEG
spectrum of M81* obtained on MJD 53944.  The data are acceptably fit
using a simple power-law function with a photon index of $\Gamma =
1.70\pm 0.07$ (shown in red).  The lower panel shows the ratio of the
observed spectrum to the power-law model.}
\medskip

In this paper, we present contemporaneous X-ray and radio observations
of M81* made using {\it Chandra} and the VLA, with visits separated by
approximately 20 days.  The observations and data reduction methods
are described in Section 2.  Our analysis and results are presented in
Section 3.  We do not find a clear correlation between radio and X-ray
emission in M81*, though a small number of simultaneous points were
obtained and span a factor of approximately two in X-ray flux.  These
results are discussed in Section 4.

\section{Observations and Data Reduction}
We observed M81* on ten occasions using {\it Chandra}.  Each
observation achieved a total exposure of approximately 15~ksec (see
Table 1).  In order to minimize photon pile-up in the zeroth order ACIS
image, the HETGS was inserted into the light path in each case.  The
ACIS chips were operated in ``FAINT'' mode.

We used CIAO version 4.0.2 in processing the {\it Chandra} data.
First-order dispersed spectra from the MEG and HEG were split from
the standard ``pha2'' file, and associated instrument response files
were constructed.  The first-order MEG spectra and responses were then
added using the CIAO tool ``add\_grating\_spectra''; the first-order HEG
spectra and responses were added in the same way.  The zeroth-order
ACIS spectra and responses were generated using the CIAO tool
``psextract''.  In each case, a circular region was used to extract
the source flux and a radially--offset annular region was used to
extract the background flux.  All spectra were grouped to require at
least 20 counts per bin using the FTOOL ``grppha'', in order to ensure
the validity of $\chi^{2}$ statistics.

The VLA also observed M81* on ten occasions (see Table 1).  Useful
data were obtained on seven occasions that coincide with the {\it
Chandra} X-ray observations.  All observations were obtained at 8.4
GHz.  The first three coincident exposures were obtained in the ``A''
configuration (achieving a typical angular resolution of approximately
0.3''), while the last four were obtained in the ``B'' configuration
(achieving a typical angular resolution of approximately 1'').
Standard compact calibrator sources were used to calibrate phase and
amplitude variations, and to set the overall amplitude scale.  The
average flux density measured in each exposure is reported in Table 1.

\centerline{~\psfig{file=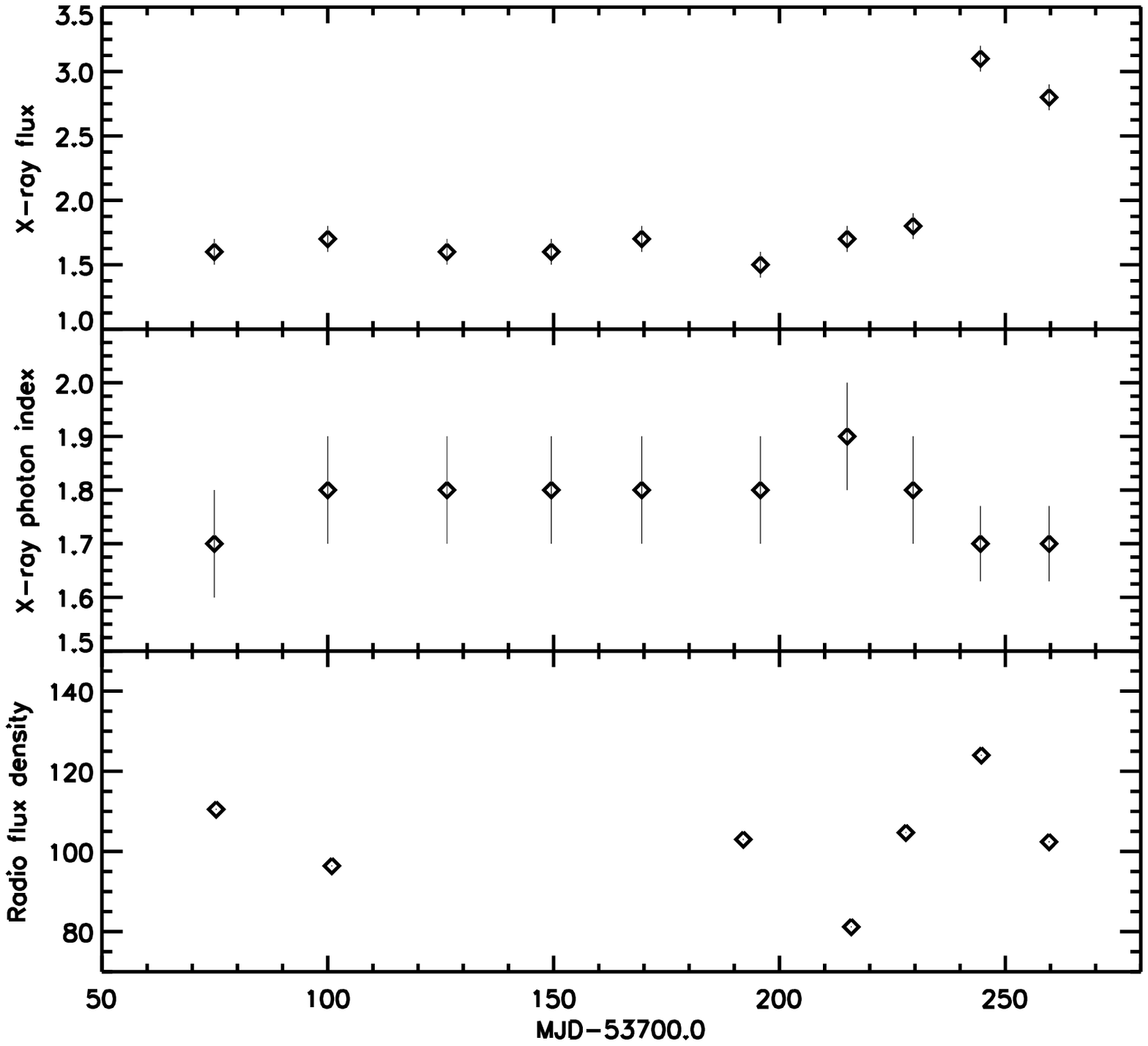,width=3.2in}~}
\figcaption[t]{\footnotesize The plot above shows the evolution of
  0.5--10.0 keV X-ray flux, X-ray photon index, and 8.4~GHz radio flux
  density observed in M81* with time.  The X-ray flux is plotted in
  units of $10^{-11}~{\rm erg}~{\rm cm}^{-2}~{\rm s}^{-1}$.  The radio
  flux density is plotted in units of 100~mJy.  The errors shown on
  all quantities are 1$\sigma$ errors; the flux density errors are
  smaller than the plotting symbols.}
\medskip

\section{Analysis and Results}
The {\it Chandra} X-ray spectra were analyzed using XSPEC version 12.4
(Arnaud 1996).  Spectral fits were made in the 0.5--10.0~keV band.  All
of the errors reported in this work are $1\sigma$ confidence errors.
In calculating luminosity values, distances were assumed to be
absolute, and uncertainties in luminosity were derived from the flux
uncertainties.

We initially made separate fits to the zeroth-order, combined MEG, and
combined HEG spectra.  In all direct fits, the equivalent neutral
hydrogen column density drifted towards zero, which is unphysical.  A
value of $4.1\times 10^{20}~{\rm cm}^{-2}$ is expected along this line
of sight (Dickey \& Lockman 1990), but this value is too low to be
constrained directly in the MEG spectra obtained.  For consistency,
then, the expected value was fixed in all fits.  All of the spectra
were acceptably fit ($\chi^{2}/\nu \leq 1.0$, where $\nu$ is the
number of degrees of freedom in the fit) with a simple power-law model
(see Figure 1).  The spectrum of M81* is likely more complex, mostly
owing to local diffuse emission (Young et al.\ 2007), but a simple
power-law is an acceptable fit to the modest spectra obtained in our
observations.

The zeroth-order spectra suffer from photon pile-up, and are not
robust.  Particularly in the last two {\it Chandra} observations,
where the flux is higher, the best-fit power-law photon index was
found to be harder.  This is consistent with multiple low-energy
X-rays being detected as single high energy photons.  Moreover, the
data/model ratio in each spectrum shows an increasing positive trend
with energy.  The HEG spectra contain many fewer photons than the MEG
spectra, and were found to be of little help in constraning the souce
flux or spectral index.  We therefore restriced our flux analysis to
the combined first-order MEG spectra.  The second spectrum listed in
Table 1, for instance, has just over 2800 photons, and the penultimate
spectrum has 4900 photons.

The results of our spectral analysis of each observation are detailed
in Table 1.  Figure 2 shows the time evolution of the X-ray flux,
X-ray power-law photon index, and radio flux density.  Between MJD
53900.0 and MJD 53950.0, the X-ray flux 

\centerline{~\psfig{file=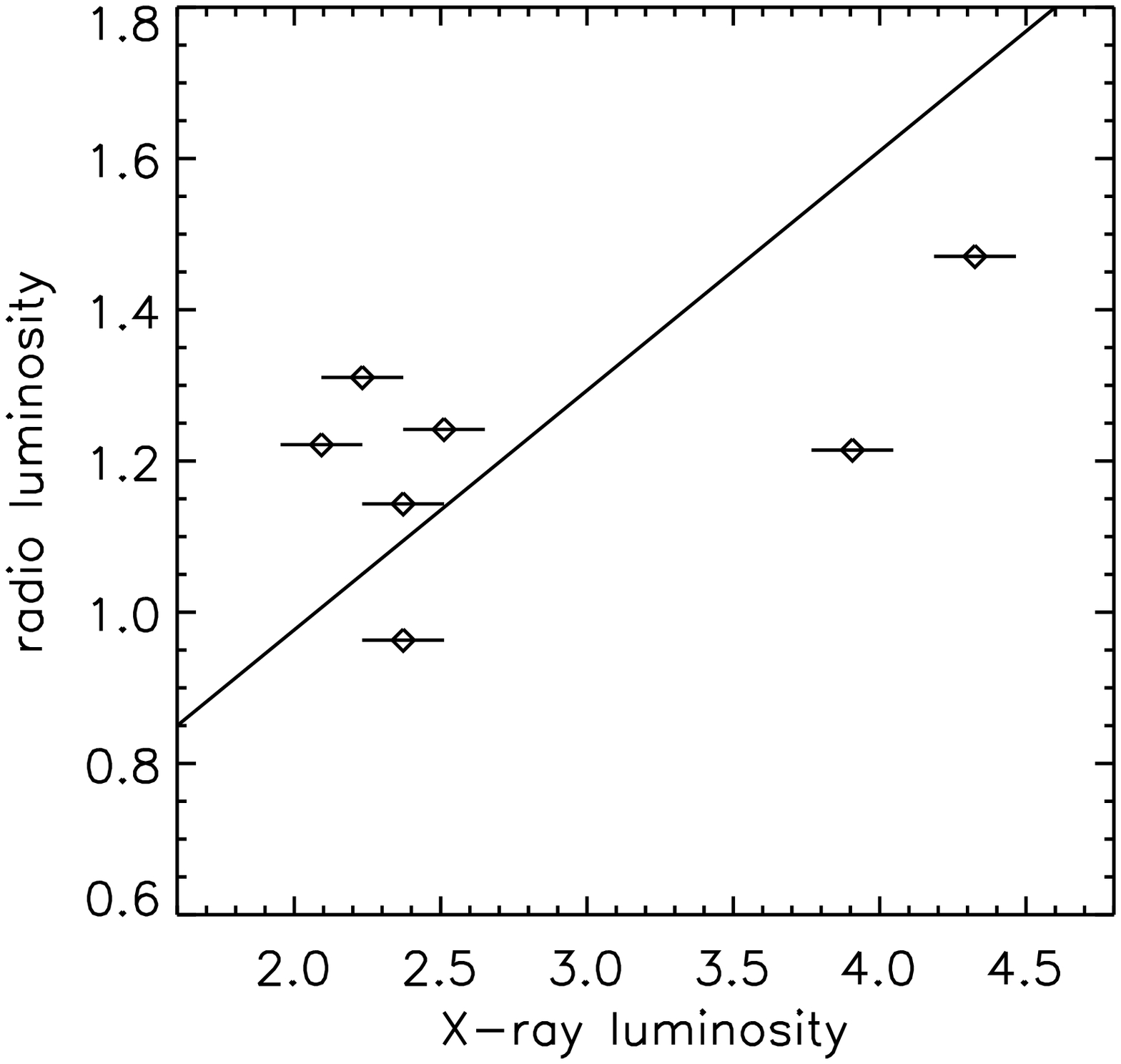,width=3.2in}~}
\figcaption[t]{\footnotesize The 8.4~GHz radio luminosity of M81* (in
units of $10^{37}$ erg/s) is plotted against its 0.3--10.0~keV X-ray
flux (in units of $10^{40}$~erg/s) in the figure above.  The errors
shown are 1$\sigma$ errors; the error on radio luminosity is very
small.  In spectrally hard black hole sources with jets, a $L_{R}
\propto L_{X}^{0.7}$ is expected.  The solid line above plots $L_{R} =
C\times L_{X}^{0.7}$, where $C=6\times 10^{8}$ was chosen arbitrarly.
The X-ray and radio luminosity values are not significantly
correlated.}
\medskip

\noindent increases by a factor of
approximately two, and the radio flux density increases by slightly
less than a factor of two.  The X-ray spectral index does not vary
significantly during the course of our observations.  In each spectrum,
the index is consistent with $\Gamma = 1.7$, which is fairly
typical of Seyferts (see, e.g., Reynolds 1997) and consistent
with prior {\it Chandra} obsevations of M81* (Young et al.\ 2007).  A
power-law is not a unique description of the data; bremsstrahlung
models also yield acceptable fits with temperatures of ${\rm kT} =
5\pm 2$~keV.

Assuming a flat radio spectrum, we calculated the radio luminosity in
the narrow VLA band centered at 8.4 GHz.  We also calculated the
unabsorbed X-ray flux in the 0.5--10.0~keV band.  We assumed a distance
of 3.6~Mpc to M81* (Freedman et al.\ 1994).  This radio luminosity is
plotted versus X-ray luminosity in Figure 3.  The Spearman's rank
correlation coefficient for the flux values is 0.23, and it is
apparent in Figure 3 that there is no strong correlation between the
X-ray and radio luminosity.  

In Figure 4, the X-ray power-law photon index is plotted versus the
X-ray luminosity.  Here again, there is no clear correlation visible
in the plot.  The Spearman's rank correlation coefficient for these
quantities is -0.39, indicating that there is no significant
correlation.

The sampling rate of the X-ray light curve is $21\pm 5$ days and
that of the radio light curve is $19\pm 5$ days.  Therefore, the
radio and X-ray peak around MJD 53944.6 are formally consistent with
being simultaneous, and any delay between radio and X-rays is
$\leq$20 days.  Given the sampling rate, it is likely that the
radio/X-ray flare is caused by a factor of $\sim$2 change in $\dot M$
between MJD 53929--53960.

\section{Discussion and Conclusions}
The nature of the inner accretion flow in LLAGN is not yet clear.  It
is possible that LLAGN retain many of the characterstics of Seyferts,
perhaps including an accretion disk extending to the ISCO
(e.g. Herrnstein et al.\ 1998, Maoz 2007).  It is also possible that
LLAGN are more like under-luminous sources, such as Sgr. A*, and best
described in terms of an ADAF or 

\centerline{~\psfig{file=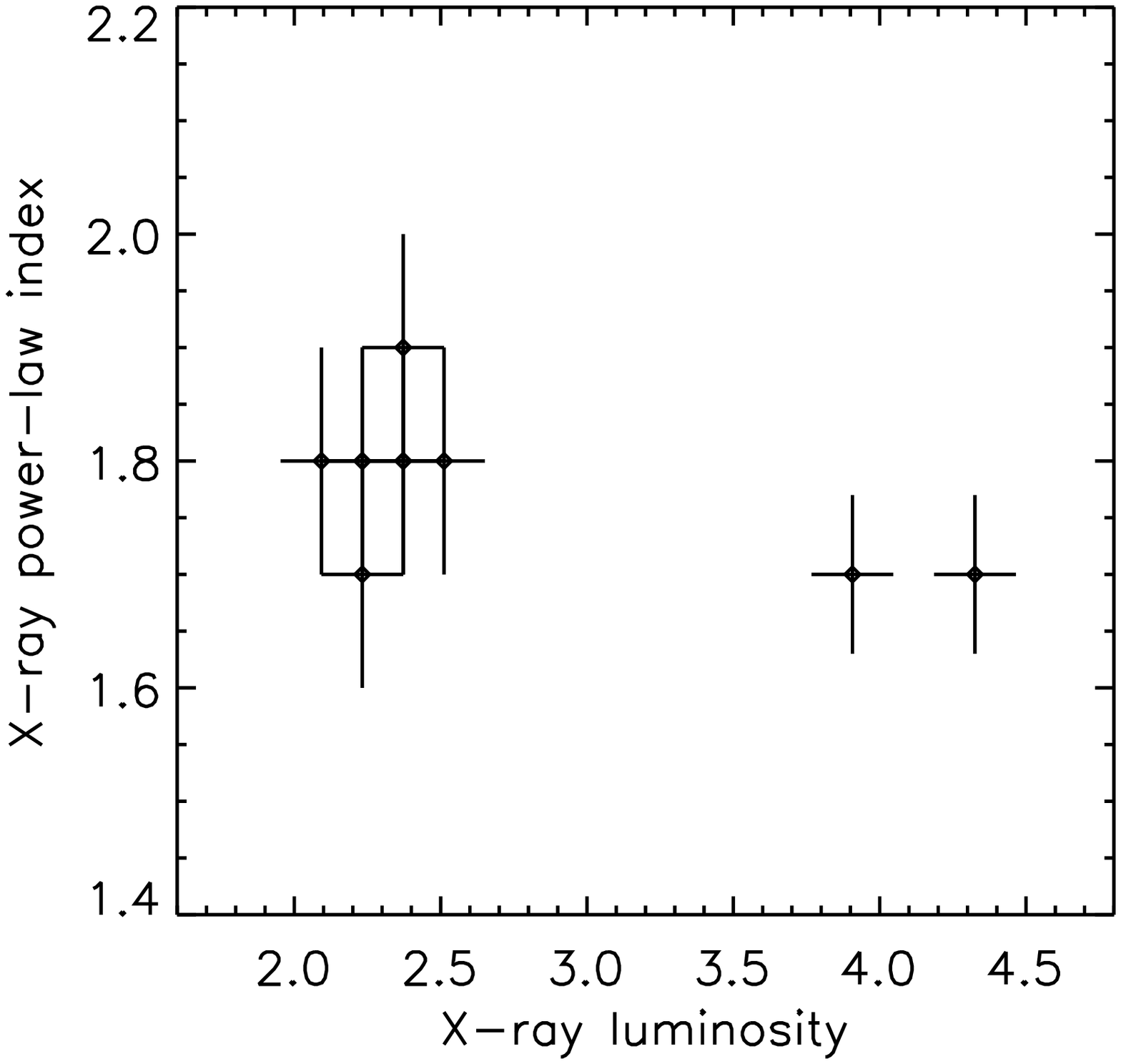,width=3.2in}~}
\figcaption[t]{\footnotesize The X-ray photon power-law index is
plotted versus the 0.3--10.0~keV X-ray luminosity of M81* (in units of
$10^{40}$~erg/s) in the figure above.  There is no evidence of the
correlations observed in Seyferts and stellar-mass black holes,
though a relatively small range in luminosity and spectral index is
sampled in our observations.}
\medskip

\noindent coupled ADAF-jet system (see, e.g., Nemmen et al.\ 2010).
In this work, we have attempted to explore the the nature of the inner
accretion flow in an LLAGN by examining evidence for a disk-jet
connection in M81*.  Based on observations of QPOs with a frequency of
$\sim 6$~Hz in stellar-mass black holes just prior to jet ejection
episodes, we sampled a commensurate timescale in M81*.  If this
timescale is an orbital period, it implies a radius that is compatible
with lower limits on the inner radius of the accretion disk in M81*
based on modeling of the Fe K$\alpha$ emission line detected in deep
obsevations of this source (${\rm R} \geq 50-60~{\rm GM}/{\rm c}^{2}$;
Dewangan et al.\ 2004, Young et al.\ 2007).  If the innermost
accretion flow is advective, so that it is geometrically thick but
retains some viscosity and angular momentum, then our sampling
timescale is also commensurate with the viscous timescale in the very
innermost region around the black hole (e.g. $6 {\rm GM}/{\rm
c}^{2}$).  Our monitoring observations improve upon many prior
investigations of disk-jet connections in many supermassive and
stellar-mass black holes in that our radio and X-ray observations were
nearly simultaneous.

At low fractions of the Eddington limit, stellar-mass black holes have
hard X-ray spectra (see, e.g., Miller et al.\ 2006, Miller, Homan, \&
Miniutti 2006, Tomsick et al.\ 2008, Reis, Fabian, \& Miller 2010; for
a review, see Remillard \& McClintock 2006) and radio emission that is
consistent with a compact jet (Fender, Homan, \& Belloni 2009).  Both
in individual sources and in an ensemble, radio and X-ray emission are
related by the expression $L_{R} \propto L_{X}^{0.7}$ (Gallo, Fender,
\& Pooley 2003; see, however, Jonker et al.\ 2009).

Our data do not strongly confirm nor reject the possibility that M81*
regulates its radio and X-ray output according to $L_{R} \propto
L_{X}^{0.7}$ (see Figure 3).  However, it is clear that if M81* does
follow this relation, the regulation of its radiation is not rigid on
short time scales.  Recent work on Sgr A* shows that the source
approaches the expected relation when it flares, but otherwise falls
below it (Markoff 2005).  These outcomes suggest that black holes
might generally channel a fixed fraction of the matter inflow (traced
by X-ray emission) into a jet (traced by radio emission), but not
necessarily at every moment.  Said differently: the energy channeled
into jets at any given time may vary as it is likely to be somewhat
stochastic, but on average the expected relationship may hold.  In
past investigations, scatter in the $L_{R} \propto L_{X}^{0.7}$
relation and the fundamental plane (e.g. Gultekin et al.\ 2009) could
plausibly be explained in terms of non-simultaneous X-ray and radio
observations; our results suggest a degree of intrinsic scatter.

Within states where compact, steady jets are produced, spectral
hardness and luminosity are positively correlated in stellar-mass
black holes (e.g. Tomsick et al.\ 2001; also see Rykoff et al.\ 2007).
Hardening of this kind has also been observed in Sgr A* (Baganoff et
al.\ 2001).  In constrast, Seyferts appear to become spectrally
softer with higher X-ray luminosity (see Vaughan \& Fabian 2004).  The
X-ray spectrum of M81* does not show a strong trend with luminosity,
and we are not able to characterize the variability of M81* as being
more like Sgr A* or more typical of Seyferts.

Recent modeling of the broad-band spectrum of M81* suggests that its
accretion flow may be very similar to that in the hard state of
stellar-mass black holes and Sgr A*.  Markoff et al.\ (2008) showed
that the same jet-dominated broad-band accretion flow model can be
applied to stellar-mass black holes, Sgr A*, and to M81*.  The
stellar-mass black hole V404 Cyg may be the source in which X-ray
observations permit the best constraints on the inner accretion flow
at $10^{-5}~{\rm L}_{\rm Edd.}$.  Recent analysis by Bradley et al.\
(2007) measure a temperature of ${\rm kT} \simeq 5$~keV with
bremsstrahlung models.  This is much too cold for even the outer parts
of a simple advection-dominated accretion flow, for which temperatures
of ${\rm kT} \geq 85$~keV are predicted (Narayan \& Yi 1995).
Similarly, fits to our spectra of M81* with bremsstrahlung models give
${\rm kT} = 5\pm 2$~keV.  In this sense, then, the X-ray spectrum of
M81* is inconsistent with very simple ADAF models because a
$\sim$5~keV plasma is too cold to be compatible with such models.

On the other hand, Young et al. (2007) showed that a combined
282\,ksec {\it Chandra} spectrum of M81* (essentially at the flux
level of the first eight observations presented here) could be
described by a model that was dominated by emission from collisional
plasma with temperatures ranging from 1--100\,keV.  In terms of an
emission measure analysis, the peak emission came from $\approx
10$--30\ keV plasma.  Similarly, one could construct an ADAF type
model where $kT \propto R^{-0.5}$, with emission ranging from $\approx
1$--$10^4\, {\rm GM}/{\rm c}^2$.  The construction of such models,
however, relied upon the detection and measurement of plasma emission
line features, which are too weak to detect in any of our short,
individual observations.  (We have verified, however, that the basic
line structure reported by Young et al. is unaltered in the full,
combined 450\,ksec spectra.)  In principle, if one could associate
line variations with the factor $> 2$ continuum flux level variations
shown here, especially for the lower temperature lines that should
arise in the less central parts of the system, this would place strong
constraints on any ADAF type model.

The large increase in X-ray flux detected in our last two observations
occurred on a time scale shorter than two weeks, which corresponds to
a light travel distance of $\approx 10^{3.5}~{\rm GM}/{\rm c}^2$.
This is smaller than the ADAF emission region postulated by Young et
al. (2007), which in any case would respond on time scales longer than
the light travel time.  Thus, any correlated line/continuum changes
would alter the simple ADAF assumptions of emission dominated by a
hot, optically thin, flow.  For example, in a situation where the hot
inner flow and thin disk partially overlap, a transition region with a
lower temperature may be expected. Transition regions have been
treated in some detail in numerous works, including Blandford \&
Begelman (1999) and Meyer, Liu, and Meyer-Hofmeister et al.\ (2000).
Emission originating in a transition region can potentially explain
our spectral and variability results, and those reported by Young et al.\
(2007).  Moreover, this possibility is qualitatively consistent with
evidence of thin disks extending to small radii at low Eddington
fractions in LLAGNs and LINERS (Maoz 2007) but still allows for a
coupled ADAF plus jet system like that described by Nemmen et al.\
(2010).

Decisive observations may be feasible with the proposed
International X-ray Observatory (IXO): a single 30~ksec observation
with IXO will achieve a sensitivity greater than that in the combined
282~ksec {\it Chandra} exposure analyzed by Young et al.\ (2007).  The
higher spectral resolution of the calorimeter expected to fly aboard
IXO will facilitate both the detection of weak lines and the detection
of small velocity shifts.  If the innermost accretion flow in M81* is
a dynamic environment where X-ray flares help to drive a jet and/or a
wind, IXO spectroscopy will be able to detect corresponding variations
in the line spectrum discussed by Young et al.\ (2007).  If a weak
iron line is produced in the inner disk, the sensitivity of IXO will
help to detect and resolve the dynamical information imprinted on any
such line.

\vspace{0.1in}
We thank the anonymous referee for thoughtful comments that improved
this paper.  J.M.M. gratefully acknowledges funding from the {\it
Chandra} Guest Observer program.  S. M. gratefully acknowledges
support from a Netherlands Organization for Scientific Research (NWO)
Vidi Fellowship.  We wish to thank the {\it Chandra} and {\it VLA}
observatory staff for executing this demanding program.


\begin{table}[t]
\begin{center}
\caption{X-ray and Radio Observations}
\vspace{0.1in}
\begin{footnotesize}
\begin{tabular}{llllllllll}
Obs. & ${\rm T}_{\rm X}$ & Exposure & $\Gamma$ & ${\rm F}_{X}$ & ${\rm T}_{\rm R}$ & Exposure & ${\rm S}_{8.4}$ & Configuration\\
 ~   & (MJD) & (ksec) & ~ & ($10^{-11}~{\rm erg}~{\rm cm}^{-2}~{\rm s}^{-1}$) & (MJD) & (ksec) & (mJy) & ~ \\
\hline
1 & 53774.9 & 15.0 & 1.7(1) & 1.6(1) & 53775.3 & 3.4  & 110.5(1) & A \\
2 & 53800.0 & 15.0 & 1.8(1) & 1.7(1) & 53800.9 & 5.3  & 96.4(1) & A\\
3 & 53826.4 & 15.0 & 1.8(1) & 1.6(1) & --      & -- & -- & --\\
4 & 53849.5 & 14.7 & 1.8(1) & 1.6(1) & --      & -- & -- & -- \\
5 & 53869.5 & 15.0 & 1.8(1) & 1.7(1) & --      & -- & -- & -- \\
6 & 53895.8 & 15.0 & 1.8(1) & 1.5(1) & 53892.0 & 5.2 & 103.0(1) & A \\
7 & 53915.0 & 14.9 & 1.9(1) & 1.7(1) & 53915.9 & 5.3 & 81.2(1) & B \\
8 & 53929.6 & 15.1 & 1.8(1) & 1.8(1) & 53928.0 & 5.2 & 104.7(1) & B \\
9 & 53944.5 & 14.6 & 1.70(7) & 3.1(1) & 53944.7 & 6.3 & 124.0(1) & B \\
10 & 53959.7 & 15.0 & 1.70(7) & 2.8(1) & 53959.7 & 5.2 & 102.4(1) & B \\
\hline
\end{tabular}
\vspace{-0.2in}
\tablecomments{The table above lists observation times and properties
of the flux observed from M81* through our joint monitoring program.
The X-ray spectra were fit with simple absorbed power-law models; the
power-law index and resulting 0.5--10.0~kV flux are reported above.
The radio flux density at 8.4~GHz is reported from each successful
observation with the VLA.  All errors are 1$\sigma$ confidence errors;
number in parentheses indicate the error in the last digit.}
\end{footnotesize}
\end{center}
\end{table}



\begin{references}

\reference{} Anderson, J. M., \& Ulvestad, J. S., 2005, ApJ, 627, 674

\reference{} Arnaud, K. A., 1996, ASPC, 101, 17

\reference{} Baganoff, F., et al., 2001, Nature, 413, 45

\reference{} Blandford, R. D., \& Begalman, M. C., 1999, MNRAS, 303, L1

\reference{} Bower, G., et al., 1996, AJ, 111, 1901

\reference{} Bradley, C., et al., 2007, ApJ, 667, 427

\reference{} Devereux, N., et al., 2003, AJ, 125, 1226

\reference{} Devereux, N., \& Shearer, A., 2007, ApJ, 671, 118

\reference{} Dewangan, G., Griffiths, R. E., Di Matteo, T., Schurch,
N. J., 2004, apJ, 607, 788

\reference{} Falcke, H., Kording, E., \& Markoff, S., 2004, A\&A, 414, 895

\reference{} Fender, R., Homan, J., \& Belloni, T., 2009, MNRAS, 396, 1370

\reference{} Ferreira, J., et al., 2006, A\&A, 447, 813

\reference{} Frank, J., King, A., \& Raine, D., 2002, in ``Accretion
Power in Astrophyiscs'', Cambridge University Press, Cambridge

\reference{} Freedman, W. L., et al., 1994, ApJ, 427, 628

\reference{} Gultekin, K., et al., 2009, ApJ, subm., arxiv:0907.3285

\reference{} Herrnstein, J. R., et al., 1998, ApJ, 497, L69

\reference{} Ho, L. C., 2008, ARA\&A, 46, 475

\reference{} Jonker, P. G., et al., 2009, MNRAS, in press

\reference{} Klein-Wolt, M., \& van der Klis, M., 2008, apJ, 675, 1407

\reference{} Livio, M., 2000, AIPC, 522, 275

\reference{} Maoz, D., 2007, MNRAS, 377, 1696

\reference{} Markoff, S., Falcke, H., \& Fender, R., 2001, A\&A, 372, L25

\reference{} Markoff, S., 2005, ApJ, 618, L103

\reference{} Markoff, S., et al., 2008, ApJ, 681, 905

\reference{} Meier, D., 2001, ApSSS, 276, 245

\reference{} Merloni, A., Heinz, S., \& Di Matteo, T., 2003, MNRAS, 345, 1057

\reference{} Meyer, F., Liu, B. F., \& Meyer-Hofmeister, E., 2000,
A\&A, 361, 175

\reference{} Miller, J. M., Ballantyne, D. R., Fabian, A. C., \&
Lewin, W. H. G. L., 2002, MNRAS, 335, 865

\reference{} Miller, J. M., Homan, J., Steeghs, D., Rupen, M.,
Hunstead, R. W., Wijnands, R., Charles, P., \& Fabian, A. C., 2006,
ApJ, 653, 525

\reference{} Miller, J. M., Homan, J., \& Miniutti, G., 2006, ApJ, 652, L113

\reference{} Miller, J. M., 2007, ARA\&A, 45, 441

\reference{} Nagar, N. M., Falcke, H., Wilson, A. S., \& Ulvestad,
J. S., 2002, A\&A, 392, 53

\reference{} Narayan, R., \& Yi, I., 1994, 428, L13

\reference{} Narayan, R., \& Yi, I., 1995, ApJ, 452, 710

\reference{} Nemmen, R. S., Storchi-Bergmann, T., Eracleous, M., \&
Yuan, F., 2010, In the Proceedings of ``Co-Evolution of Central Black
Holes and Galaxies'', IAU Symposium 267, eds. B. M. Peterson,
R. s. Somerville, T. Storchi-Bergmann, arxiv:1001.3174

\reference{} Nespoli, E., et al., 2003, A\&A, 413, 235

\reference{} Reis, R. C., Fabian, A. C., \& Miller, J. M., 2010, MNRAS, 402, 836

\reference{} Reynolds, C. S., 1997, MNRAS, 286, 513

\reference{} Reynolds, C. S., Garofalo, D., \& Begelman, M. C., 2006,
ApJ, 651, 1023

\reference{} Reynolds, C. S., et al., 2009, ApJ, 691, 1159

\reference{} Russell, D. M., Maitra, D., Dunn, R. J. H., \& Markoff,
S., 2010, MNRAS, in press, arxiv:1002.3729

\reference{} Tomsick, J., et al., 2008, ApJ, 680, 593

\reference{} Young, A. J., Nowak, M., Markoff, S., Marshall, H. L., \&
Canizares, C. R., 2007, ApJ, 669, 830




\end{references}
\end{document}